%% file: main.tex
\documentclass[a4paper]{article}

\usepackage{INTERSPEECH2020}
\usepackage{multirow}
\usepackage{algorithm}
\usepackage{algorithmic}
\usepackage{cite}

\title{Investigating Robustness of Adversarial Samples Detection for \\Automatic Speaker Verification\thanks{This work was done when Xu Li was an intern at Tencent AI Lab.}}

\name{Xu Li$^{1}$, Na Li$^2$, Jinghua Zhong$^{3}$, Xixin Wu$^{4}$, Xunying Liu$^1$, Dan Su$^2$, Dong Yu$^2$, Helen Meng$^1$}
\address{
  $^1$The Chinese University of Hong Kong, Hong Kong SAR, China\\
  $^2$Tencent AI Lab, Tencent, Shenzhen, China\\
  $^3$SpeechX limited, Shenzhen, China\\
  $^4$Department of Engineering, University of Cambridge, UK\\
  {\small \tt \{xuli, wuxx, xyliu, hmmeng\}@se.cuhk.edu.hk, lina011779@126.com, jhzhong@speechx.cn, \{dansu, dyu\}@tencent.com}}
  
\email{}

\begin{document}
\ninept

\maketitle
\input{abstract}
\input{introduction}
\input{adv-samples-gen}
\input{adv-samples-det}
\input{experiment-results}

\input{conclusion}
\input{acknowledgements}

\bibliographystyle{IEEEtran}

\bibliography{mybib}


\end{document}

%% file: abstract.tex
\begin{abstract}
Recently adversarial attacks on automatic speaker verification (ASV) systems attracted widespread attention as they pose severe threats to ASV systems. However, methods to defend against such attacks are limited. Existing approaches mainly focus on retraining ASV systems with adversarial data augmentation. Also, countermeasure robustness against different attack settings are insufficiently investigated. Orthogonal to prior approaches, this work proposes to defend ASV systems against adversarial attacks with a separate detection network, rather than augmenting adversarial data into ASV training. A VGG-like binary classification detector is introduced and demonstrated to be effective on detecting adversarial samples. To investigate detector robustness in a realistic defense scenario where unseen attack settings 
may exist, we analyze various kinds of unseen attack settings' impact and observe 
that the detector is robust (6.27\% $EER_{det}$ degradation in the worst case) against unseen substitute ASV systems, but it has weak robustness (50.37\% $EER_{det}$ degradation in the worst case) against unseen perturbation methods. The weak robustness against unseen perturbation methods shows a direction for developing stronger countermeasures.

\end{abstract}

%% file: introduction.tex
\section{Introduction}
\label{sec:intro}
Automatic speaker verification (ASV) systems aim at confirming a claimed speaker identity against a spoken utterance. It has been widely applied into commercial devices and authorization tools.
However, recent studies have shown that a well-trained ASV system could be deceived by malicious attacks\cite{kinnunen2012vulnerability,shchemelinin2014vulnerability,wu2015spoofing}. In the last decade, the speaker verification community held several ASVspoof challenge competitions\cite{wu2015asvspoof,kinnunen2017asvspoof,todisco2019asvspoof} to develop countermeasures mainly against replay \cite{williams2019speech,cai2019dku}, speech synthesis \cite{hanilcci2015classifiers,wu2012detecting} and voice conversion \cite{wu2012detecting,correia2014preventing} attacks. 

Very recently, another threat, named adversarial attacks, has been explored on ASV systems. Adversarial attacks slightly perturb the input so that the system will make incorrect decisions. Kreuk et al. \cite{kreuk2018fooling} added adversarial perturbations into a testing utterance to attack an end-to-end ASV system. The attack was verified to be successful in both cross-feature and cross-corpus settings. Li et al. \cite{li2020adversarial} extended the studies into other ASV frameworks and observed the adversarial transferability from one ASV to attack another ASV. Also some works explored adversarial attacks in practical real-time scenarios \cite{chen2019real,li2020practical,xie2020real} and attacks on spoofing countermeasures \cite{liu2019adversarial}.

Apart from the effective perturbations that pose severe threats on ASV systems, the perturbation variations caused by different attack settings also bring difficulty in developing defense approaches. In a realistic attack, different substitute ASV systems can be used to craft adversarial samples and perform effective attack on the target ASV system in a transferable way \cite{li2020adversarial}. The choice of a substitute ASV system, as one of attack settings, results in different perturbation patterns.
Besides, perturbation patterns also vary greatly across perturbation methods \cite{kurakin2016adversarial} with different perturbation configurations, e.g. perturbation degrees. So countermeasure robustness against different attack settings, including substitute ASV systems, perturbation methods along with perturbation configurations, is another important concern.

Defense approaches against adversarial attacks have been investigated mostly in the image domain \cite{song2017pixeldefend,xu2017feature,gong2017adversarial}. Defense approaches explored in ASV area are still very limited. Wang et al. \cite{wang2019adversarial} leveraged adversarial samples into training an end-to-end ASV as a regularization to improve system robustness. Wu et al. \cite{wu2020defense} adopted a combination of spatial smoothing \cite{xu2017feature} and adversarial training \cite{goodfellow2014explaining} to strengthen countermeasures against adversarial samples. Both methods are found to be effective. However, they need to retrain a well-developed ASV system with adversarial data augmentation. To the best of our knowledge, no existing work investigates countermeasure robustness against different attack settings of spoofing ASV systems.

Inspired by \cite{gong2017adversarial,samizade2019adversarial}, this work makes the first attempt to defend ASV systems against adversarial attacks with a separate detection network. A VGG-like \cite{nagrani2020voxceleb} binary classification system is introduced to capture the difference between adversarial and genuine samples, and predict whether an input is adversarial or not. A separate detection countermeasure has the following advantages: 1) It separates the defense part and speaker verification into two independent stages, which avoids retraining a well-developed ASV model. 2) Since most existing countermeasures for replay and synthetic speech attacks are based on a separate detection network \cite{williams2019speech,cai2019dku,hanilcci2015classifiers}, the proposed approach provides the feasibility to develop a unified countermeasure against all spoofing attacks.

In a realistic defense scenario, attack settings cannot be accessed by the defender so that the proposed detector can be degraded by unseen attack settings.
To investigate detector robustness in such a realistic scenario and provide directions for developing stronger countermeasures, this work also gives a robustness discussion based on unseen attack settings, including substitute ASV systems, perturbation methods and perturbation degrees.
In this work, the three most representative ASV frameworks are used as variations: Gaussian mixture model (GMM) i-vector system \cite{dehak2010front}, time delay neural network (TDNN) x-vector system \cite{snyder2018x} and ResNet-34 r-vector system \cite{zeinali2019but}. Two of the most effective perturbation methods, i.e. basic iterative method (BIM) \cite{kurakin2016adversarial} and Jacobian-based saliency map approach (JSMA) \cite{papernot2016limitations}, are applied to generate adversarial samples. 

The contributions of this work include: 1) Design of a dedicated defense network against adversarial attacks, rather than augmenting adversarial samples into ASV training; 2) Introduction of a VGG-like network and demonstrating its effectiveness on detecting adversarial samples; 3) Investigation of detector robustness against unseen attack settings to uncover vulnerability and lack of robustness against unseen perturbation methods, which provides directions for developing stronger countermeasures.

The remaining of this paper is organized as follows: Section~\ref{sec:adv-samples-gen} details the process of adversarial samples generation. The proposed adversarial samples detection network is illustrated in Section~\ref{sec:adv-samples-det}. Section~\ref{sec:expt-results} analyzes the experiment results. Finally, Section~\ref{sec:conclusion} summaries this work.

%% file: adv-samples-gen.tex
\section{Adversarial Samples Generation}
\label{sec:adv-samples-gen}
In a speaker verification task, given acoustic features of the enrollment utterance $\boldsymbol{X^{(e)}}$ and testing utterance $\boldsymbol{X^{(t)}}$, a well-trained system function $S$ with parameters $\boldsymbol{\theta}$ will predict a similarity score, which indicates speaker similarity between the enrollment and testing utterances. 
In a realistic scenario, the owner's enrollment utterance $\boldsymbol{X^{(e)}}$ is implicitly embedded within the ASV system, while $\boldsymbol{X^{(t)}}$ is provided by the customer for identity confirmation.
From an adversary's perspective, it will optimize a perturbation $\boldsymbol{\delta_X}$ to be added on $\boldsymbol{X^{(t)}}$ so that the system will behave incorrectly: either falsely rejecting the true target's voice or falsely accepting the imposter's voice. The optimization problem can be formulated as Eq.~\ref{eq:opt-formulation} and \ref{eq:opt-formulation-k}:
\begin{align}
    &\boldsymbol{\delta_X} = \arg \max_{\Vert \boldsymbol{\delta_X} \Vert_p \leq \epsilon} k \times S_{\boldsymbol{\theta}}(\boldsymbol{X^{(e)}}, \boldsymbol{X^{(t)}}+\boldsymbol{\delta_X}) \label{eq:opt-formulation} \\
    &k = \{ \begin{array}{cc}
     -1, & \text{target trial} \\
     1, & \text{non-target trial}
    \end{array}
    \label{eq:opt-formulation-k}
\end{align}
where the constraint $p$-norm of $\boldsymbol{\delta_X}$ within perturbation degree $\epsilon$ guarantees a subtle perturbation so that human cannot perceive the difference between adversarial and genuine samples.

We leverage three different ASV system architectures and two perturbation methods in our experiments. The details for ASV systems and perturbation methods are illustrated in Sections~\ref{sec:asv-sys} and \ref{sec:perturbation-method}, respectively.

\subsection{ASV systems}
\label{sec:asv-sys}
Three ASV systems involved are as follows: GMM i-vector with probabilistic linear discriminant analysis (PLDA) back-end \cite{dehak2010front}, TDNN x-vector with PLDA back-end \cite{snyder2018x} and ResNet-34 r-vector with cosine back-end \cite{zeinali2019but}. All systems adopt cepstral frequency with configurations in \cite{li2020adversarial} as input.

The i-vector system \cite{dehak2010front} consists of 2048 mixtures with full covariance matrix. $T$ matrix projects utterance statistics into a 400-dimension i-vector space. The i-vectors are centered and length-normalized before PLDA modeling.

The x-vector system is configured as \cite{snyder2018x}, except that additive angular margin (AAM)-softmax loss \cite{xiang2019margin} with hyper-parameters \{$m=0.3$, $s=32$\} is used for training. Extracted x-vectors are centered and projected by a 200-dimension LDA, then length-normalized before PLDA modeling.

The r-vector system has the same architecture as \cite{zeinali2019but}, and AAM-softmax loss \cite{xiang2019margin} with hyper-parameters \{$m=0.2$, $s=30$\} is used for training networks. Extracted r-vectors are centered and length-normalized before cosine scoring.

\subsection{Perturbation methods}
\label{sec:perturbation-method}

BIM perturbs the genuine input $\boldsymbol{X^{(t)}}$ towards the gradient of the objective w.r.t. $\boldsymbol{X^{(t)}}$ in a multiple-step manner. It optimizes the perturbation with the norm constraint parameter $p$ in Eq.~\ref{eq:opt-formulation} being $\infty$. Starting from the genuine input $\boldsymbol{X^{(t)}_0} = \boldsymbol{X^{(t)}}$, the input is perturbed iteratively as follows: 
\begin{align}
   \nonumber
    \boldsymbol{X^{(t)}_{n+1}} = clip_{\boldsymbol{X^{(t)}}, \epsilon}(\boldsymbol{X^{(t)}_{n}}+\alpha si&gn(\nabla_{\boldsymbol{X^{(t)}_{n}}} S_{\boldsymbol{\theta}}(\boldsymbol{X^{(e)}}, \boldsymbol{X^{(t)}_{n}}))), \\
    &\text{for $n = 0, ..., N-1$} \label{eq:bim-solu}
\end{align}
where $sign$ is a function that takes the sign of the gradient, $\alpha$ absorbs the trial indicator $k$ in Eq.~\ref{eq:opt-formulation-k} and its absolute value is the step size, $N$ is the number of iterations and $clip_{\boldsymbol{X^{(t)}}, \epsilon}(\boldsymbol{X})$ holds the norm constraints by applying element-wise clipping such that $\Vert \boldsymbol{X}-\boldsymbol{X^{(t)}} \Vert_{\infty} \leq \epsilon$. In our experiments, $N$ is set as 5, and $\alpha$ is set as perturbation degree devided by $N$.



JSMA is another effective perturbation method to craft adversarial samples. Unlike BIM that adds perturbations to the whole input, JSMA perturbs only one bit at a time. In each iteration, it selects the bit with the most significant effects on output to be perturbed. 
With this purpose, a saliency score is calculated for each bit and bit with the highest score is chosen to be perturbed. We formulate the algorithm specialized in our case, as shown in Algorithm~\ref{alg:jsma-method}. The $saliency\_map$ at Step~\ref{code:jsma-5} computes the absolute value of gradient $\boldsymbol{G}$ while masking out the bits already reach the constraint boundary: $saliency\_map(\boldsymbol{G}, \boldsymbol{\Gamma})= \boldsymbol{G}^{abs} \odot \boldsymbol{\Gamma}$, where $\boldsymbol{G}^{abs}$ is the element-wise absolute value of $\boldsymbol{G}$ and $\odot$ is an element-wise product operator. In this work, $N$ is set as 300 iterations, and $\alpha$ is set as half of the perturbation degree.

\begin{algorithm}
\caption{JSMA perturbation method\\
$\boldsymbol{X^{(e)}}$ and $\boldsymbol{X^{(t)}}$ are acoustic features of enrollment and testing utterances, respectively. $S_{\boldsymbol{\theta}}$ is the system function with parameters, $\alpha$ is the step size, $\epsilon$ is the perturbation degree, and $N$ is the number of iterations. $\boldsymbol{\Gamma}$ is a mask matrix having the same size with $\boldsymbol{X^{(t)}}$, initialized with all-one element matrix $\boldsymbol{E}$.}
\label{alg:jsma-method}
\begin{algorithmic}[1]
\REQUIRE $\boldsymbol{X^{(e)}}$, $\boldsymbol{X^{(t)}}$, $S_{\boldsymbol{\theta}}$, $\alpha$, $\epsilon$, $N$
\STATE $\boldsymbol{X^{(t)}_{adv}}=\boldsymbol{X^{(t)}}$, $\boldsymbol{\Gamma} = \boldsymbol{E}$, $\boldsymbol{\delta_{X}}=\boldsymbol{0}$
\label{code:jsma-2}
\FOR{$i \in [1,N]$}
\label{code:jsma-3}
\STATE $\boldsymbol{G} = \nabla_{\boldsymbol{X^{(t)}_{adv}}} S_{\boldsymbol{\theta}}(\boldsymbol{X^{(e)}}, \boldsymbol{X^{(t)}_{adv}})$
\label{code:jsma-4}
\STATE $\boldsymbol{M} = saliency\_map(\boldsymbol{G}, \boldsymbol{\Gamma})$
\label{code:jsma-5}
\STATE $k_{max} = \arg \max_{k} \boldsymbol{M}_k$
\label{code:jsma-6}
\STATE $\boldsymbol{\delta_{X}}[k_{max}] = clip_{\boldsymbol{0},\epsilon} (\boldsymbol{\delta_{X}}[k_{max}]+\alpha \times sign(\boldsymbol{G}_{k_{max}}))$
\label{code:jsma-7}
\IF{$| \boldsymbol{\delta_{X}}[k_{max}] | \geq \epsilon$}
\label{code:jsma-8}
\STATE $\boldsymbol{\Gamma}_{k_{max}} = 0$
\label{code:jsma-9}
\ENDIF
\label{code:jsma-10}
\STATE $\boldsymbol{X^{(t)}_{adv}} = \boldsymbol{X^{(t)}}+\boldsymbol{\delta_{X}}$
\label{code:jsma-11}
\ENDFOR
\label{code:jsma-12}
\RETURN $\boldsymbol{X^{(t)}_{adv}}$
\label{code:jsma-13}
\end{algorithmic}
\end{algorithm}

\subsection{Dataset generation}
Our experiments are conducted on the Voxceleb1 \cite{nagrani2017voxceleb} dataset, which consists of short clips of human speech. There are in total 148,642 utterances from 1,251 speakers. Following data partitioning in \cite{nagrani2017voxceleb}, 148,642 utterances from 1211 speakers are used to train the ASV systems, and the remaining 4,874 utterances from 40 speakers are used for testing the systems and generating adversarial samples. The corpus \cite{nagrani2017voxceleb} provides totally 37,720 trials consisting of enrollment-testing utterance pairs selected from the testing utterances.

In this work, we generate adversarial samples according to the attack configuration, including the substitute ASV system, perturbation method and perturbation degree. To make a balanced dataset, for each genuine utterance, we randomly select one trial where that utterance is used to generate an adversarial counterpart. There are around 9K utterances in such an ``adversary-genuineness'' dataset, including around 4.5K adversarial and 4.5K genuine utterances.

For each specific attack configuration, we generate one ``adversary-genuineness'' dataset for training and evaluating our detection network. We separate the ``adversary-genuineness'' dataset into training and testing subsets, with 30 speaker's data for training and 10 speaker's data for testing. The speaker partitioning for training and testing is consistent among all attack configurations. This guarantees that source utterances (either a genuine utterance or an adversarial utterance generated from it) in the testing subsets cannot be observed during training.



%% file: adv-samples-det.tex
\section{Adversarial Samples Detection}
\label{sec:adv-samples-det}


In this section, we present our proposed system to detect adversarial samples. One possible feature engineering is to use the same features adopted by the protected ASV system. This provides high resistance to the most severe attack scenario where the attacker can access the whole parameters of ASV and directly add perturbations on the features adopted by ASV. With this consideration, we adopt Mel-frequency cepstral coefficients (MFCCs) as the input feature forwarded to our detection network. A pre-emphasis with coefficient of 0.97 is adopted. 25ms ``Hamming" window with step-size of 10ms is applied to extract a frame, and finally 24 cepstral coefficients are kept.

We notice some issues about adversarial samples: 1) The deviation between adversarial and genuine samples is subtle and localized on feature maps, and we shall adopt convolutional layers at bottom to effectively capture such deviations; 2) The adversarial characteristics exist in the whole utterance, so a pooling layer can be adopted to aggregate the utterance statistics for decision.
Based on these considerations, we introduce a VGG-like network structure \cite{nagrani2020voxceleb} to detect adversarial samples. The detailed architecture configurations are illustrated in Table~\ref{tab:det-config}. 4 convolutional layers at bottom to capture local feature patterns. A statistics pooling layer aggregates the mean and deviation from the last convolutional layer outputs, and forwards them to dense layers. Finally, 2 dense layers project statistics into a 2-dimensional output space for decision. The network is trained with the Adam \cite{kingma2014adam} optimizer, along with the initial learning rate as 0.001.


%% file: experiment-results.tex
\section{Experiments}
\label{sec:expt-results}

\subsection{Evaluation metrics}
\label{sec:evaluation-metrics}
To verify the effectiveness of adversarial attacks, we evaluate the ASV system performance before and under adversarial attacks in terms of equal error rate (EER) and minimum detection cost function with target trial prior to be 0.01 and 0.001, i.e. $DCF_{0.01}$ and $DCF_{0.001}$. When evaluating the detector performance, we report the detection accuracy (DA) over the ``adversary-genuineness'' testing subset. Also, regardless of the operating point, we use the detector's log softmax output at the adversarial bit as the adversarial score, and compute an EER ($EER_{det}$) over the testing subset.

\begin{table}[t]
\caption{Detailed configurations of the proposed detector.}
\label{tab:det-config}
\centering
\begin{tabular}{cccc}
         \hline
         \hline
         Layer & Structure & Activation  \\
         \hline
         Conv2D & $[2 \times 2, 64] \times 4$ & ReLU \\
         Statistics Pooling & -             & -    \\
         Flatten            & -             & -    \\
         Dense1             & 512, dropout 0.2           & ReLU \\
         Dense2             & 512, dropout 0.2           & ReLU \\
         Output             & 2             & Softmax \\
         \hline
         \hline
\end{tabular}
\end{table}

\subsection{Adversarial attack performance}
\label{sec:adv-attack-performance}

The attack results on the x-vector system are shown in Table~\ref{tab:x-vec-performance}. The results on the i-vector and r-vector systems have similar trends. From Table~\ref{tab:x-vec-performance}, we observe that the ASV system performance seriously drops when being attacked by both perturbation methods. Also, the attack effectiveness increases as the perturbation degree increases. However, the perturbations with a higher degree are easier to be detected, which will be discussed in Section~\ref{sec:robust-perturbation-degree}. This suggests a trade-off for attackers to design an effective but cannot be easily detected perturbations.

\begin{table}[h]
\caption{The x-vector system performance under different attack configurations.}
\label{tab:x-vec-performance}
\centering
\begin{tabular}{c|c|c|c|c}
\hline
\hline
\multicolumn{2}{c|}{}   & EER (\%) & $DCF_{0.01}$ & $DCF_{0.001}$ \\
\hline
\multicolumn{2}{c|}{genuine}                           &  5.97    & 0.515        & 0.695 \\ \hline
\multirow{3}{*}{BIM}                 &  $\epsilon=0.3$                  & 39.87    & 0.995        & 0.996     \\ \cline{2-5}
                                     &  $\epsilon=1.0$                  & 95.02    & 1            & 1     \\ \cline{2-5}
                                     &  $\epsilon=2.0$                  & 99.96    & 1            & 1     \\
\hline
\multirow{3}{*}{JSMA}                &  $\epsilon=1.0$                  & 20.41    & 0.880        & 0.932     \\ \cline{2-5}
                                     &  $\epsilon=3.0$                  & 48.28    & 0.995        & 0.995     \\ \cline{2-5}
                                     &  $\epsilon=5.0$                  & 60.22    & 1            & 1     \\
\hline
\hline
\end{tabular}
\end{table}

\begin{table}[b]
\caption{Detection accuracy (\%) against perturbation degrees}
\label{tab:DA-perturbation-degree}
\centering
\begin{tabular}{c|c|c|c|c}
\hline
\hline
\multicolumn{2}{c|}{\multirow{2}{*}{BIM-xvec}} & \multicolumn{3}{c}{training} \\ \cline{3-5}
\multicolumn{2}{c|}{}                           & $\epsilon=0.3$ & $\epsilon=1.0$ & $\epsilon=2.0$ \\ \hline
      \multirow{3}{*}{evaluation} & $\epsilon=0.3$ & 99.83      & 48.65      & 48.61      \\ \cline{2-5}
                                  & $\epsilon=1.0$ & 99.82      & 100.00     & 87.01      \\ \cline{2-5}
                                  & $\epsilon=2.0$ & 99.83      & 100.00     & 100.00     \\ 
\hline
\hline
\multicolumn{2}{c|}{\multirow{2}{*}{JSMA-xvec}} & \multicolumn{3}{c}{training} \\ \cline{3-5}
\multicolumn{2}{c|}{}                           & $\epsilon=1.0$ & $\epsilon=3.0$ & $\epsilon=5.0$ \\ \hline
      \multirow{3}{*}{evaluation} & $\epsilon=1.0$ & 99.44      & 59.84      & 48.61      \\ \cline{2-5}
                                  & $\epsilon=3.0$ & 99.83      & 100.00     & 98.41      \\ \cline{2-5}
                                  & $\epsilon=5.0$ & 99.83      & 100.00     & 100.00     \\
\hline
\hline
\end{tabular}
\end{table}

\subsection{Robustness against perturbation degree}
\label{sec:robust-perturbation-degree}

In this section, we discuss detector robustness against perturbation degree. Adversarial samples crafted from the x-vector system along with BIM and JSMA perturbation methods are involved. The system detection accuracy (DA) under different conditions is shown in Table~\ref{tab:DA-perturbation-degree}. The diagonal results are based on in-domain evaluation, which reflects our proposed detection network is effective and can distinguish the adversarial and genuine data with an accuracy over 99\%. It is also observed that the detector can generalize well from adversarial samples with a small perturbation to a larger perturbation. However, the performance drops greatly in the reverse direction. This indicates that we should craft small perturbations to develop our detector, so that it could defend ASV systems against adversarial samples with equal or higher degrees very well. 

\subsection{Robustness against substitute ASV systems}
\label{sec:robust-asv-sys}
In this section, we investigate detector robustness against substitute ASV systems. We conduct experiments on the BIM perturbation method with $\epsilon=0.3$ attacking i-vector, x-vector and r-vector systems. The experiment results based on JSMA method have similar observations. The DA and $EER_{det}$ are shown in Table~\ref{tab:DP-ASV-sys}. From the experimental results, we observe that 
i-vector and x-vector systems generalize well to each other but performance decreases when generalizing to r-vector system. One possible explanation is that both i-vector and x-vector systems use PLDA back-end, while r-vector system uses cosine back-end. The choice of back-end modelling may have a larger influence on perturbation patterns than utterance embedding modelling. Besides, r-vector system has better generalization compared with i-vector and x-vector systems, it perhaps implies cosine back-end has better generalization ability.

\begin{figure}[t]
    \centering
    \includegraphics[width=0.45\textwidth]{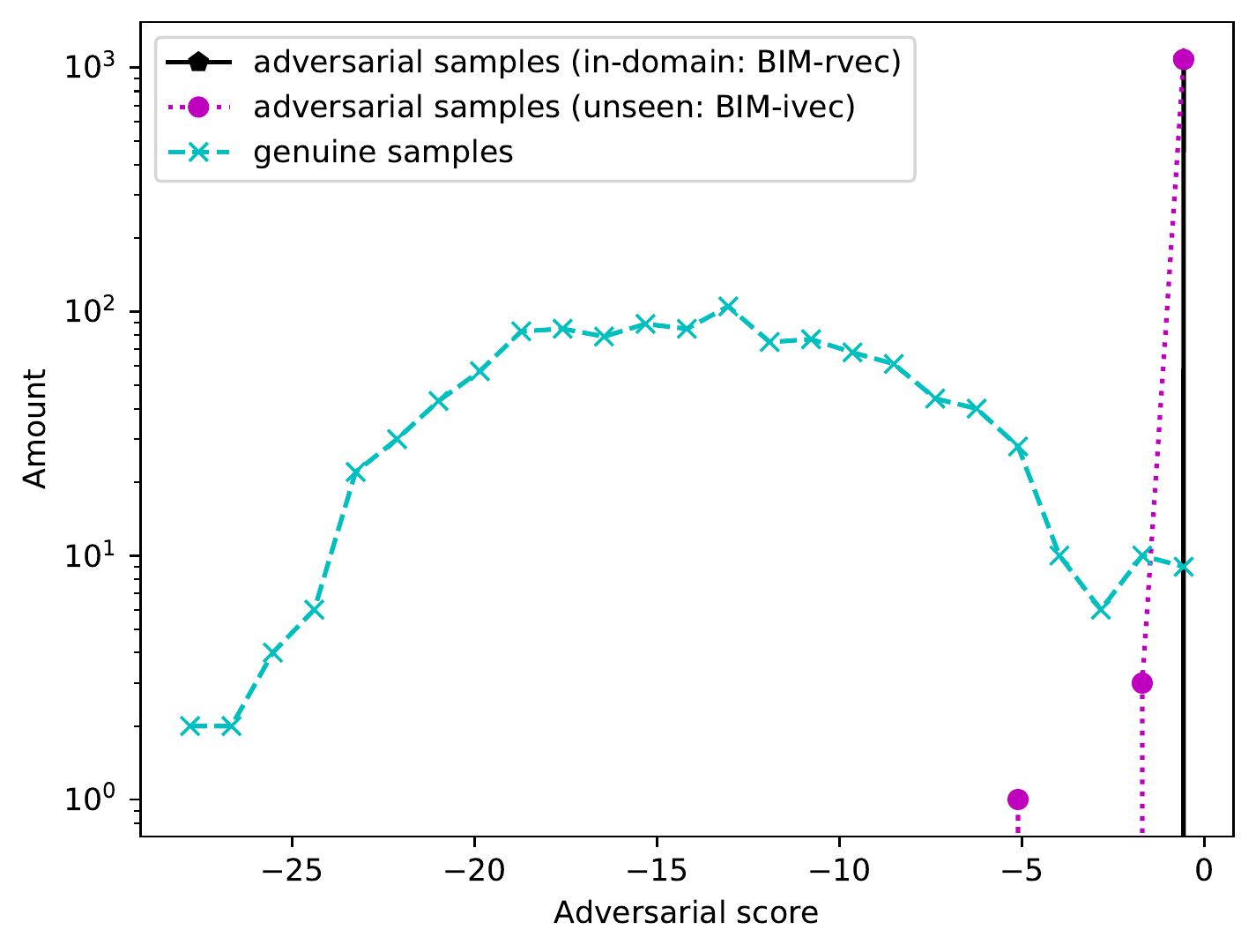}
    \caption{The adversarial score distribution for genuine samples, and adversarial samples crafted from in-domain and unseen ASV systems (training: BIM-rvec, evaluation: BIM-ivec).}
    \label{fig:score-distr-pgdrvec-pgdivec}
\end{figure}

To see the detector's ability to recognize adversarial samples, we visualize adversarial score distributions for genuine samples, in-domain and unseen adversarial samples,
as shown in Fig.~\ref{fig:score-distr-pgdrvec-pgdivec}.
It shows the detector can generalize well to unseen ASV systems by assigning high adversarial scores to most of adversarial samples. For some cases where a low DA occurs, e.g. training on i-vector while evaluated on r-vector system (72.45\%), the detector still achieves an acceptable $EER_{det}$ (6.27\%). This indicates the detector still works well but there needs a shifted operating point to detect adversarial samples.

\begin{table}[t]
\caption{Detector performance against ASV systems}
\label{tab:DP-ASV-sys}
\centering
\begin{tabular}{c|c|c|c|c}
\hline
\hline
\multicolumn{2}{c|}{\multirow{2}{*}{DA (\%)}} & \multicolumn{3}{c}{training} \\ \cline{3-5}
\multicolumn{2}{c|}{}                           & BIM-ivec & BIM-xvec & BIM-rvec \\ \hline
      \multirow{3}{*}{evaluation} & BIM-ivec & 99.87      & 99.78      & 99.44      \\ \cline{2-5}
                                  & BIM-xvec & 99.65      & 99.83     & 99.39      \\ \cline{2-5}
                                  & BIM-rvec & 72.45      & 76.38     & 99.70     \\ 
\hline
\hline
\multicolumn{2}{c|}{\multirow{2}{*}{$EER_{det}$ (\%)}} & \multicolumn{3}{c}{training} \\ \cline{3-5}
\multicolumn{2}{c|}{}                           & BIM-ivec & BIM-xvec & BIM-rvec \\ \hline
      \multirow{3}{*}{evaluation} & BIM-ivec & 0      & 0.18      & 0.55      \\ \cline{2-5}
                                  & BIM-xvec & 0.46      & 0.18     & 0.65      \\ \cline{2-5}
                                  & BIM-rvec & 6.27      & 5.90     & 0.28     \\ 
\hline
\hline
\end{tabular}
\end{table}


\subsection{Robustness against perturbation methods}

In this section, we investigate detector robustness against perturbation methods. Detector performance is evaluated by adversarial samples crafted from BIM and JSMA attacking on the x-vector system, as shown in Table~\ref{tab:DP-perturbation-method}. We observe a generalizability of 10.15\% $EER_{det}$ from JSMA to BIM and 50.55\% $EER_{det}$ from BIM to JSMA. This indicates the generalizability is not symmetric and can drop greatly in some cases to be a random guess (50.55\% $EER_{det}$). This phenomenon shows a limited detector robustness against unseen perturbation methods. The detector trained on a combination of both methods can perform well on both, which suggests that we could enlarge our training dataset to include as many existing perturbation methods as possible to enhance our model's robustness. To deal with unseen perturbation methods, we believe that a proper combination of observed perturbation methods can reinforce the detector's robustness. We leave this to future studies.


\begin{table}[ht]
\caption{Detector performance against perturbation methods}
\label{tab:DP-perturbation-method}
\centering
\begin{tabular}{c|c|c|c|c}
\hline
\hline
\multicolumn{2}{c|}{\multirow{2}{*}{DA (\%)}} & \multicolumn{3}{c}{training} \\ \cline{3-5}
\multicolumn{2}{c|}{}                    & BIM       & JSMA     & combined \\ \hline
      \multirow{2}{*}{evaluation} & BIM  & 99.83   & 57.73  &  99.48 \\ \cline{2-5}
                                  & JSMA & 48.61   & 99.44  &  99.09 \\
\hline
\hline
\multicolumn{2}{c|}{\multirow{2}{*}{$EER_{det}$ (\%)}} & \multicolumn{3}{c}{training} \\ \cline{3-5}
\multicolumn{2}{c|}{}                    & BIM       & JSMA     & combined \\ \hline
      \multirow{2}{*}{evaluation} & BIM  & 0.18   & 10.15  &  0.46 \\ \cline{2-5}
                                  & JSMA & 50.55   & 0.46  &  0.92 \\
\hline
\hline
\end{tabular}
\end{table}

%% file: conclusion.tex
\section{Conclusion}
\label{sec:conclusion}
This work proposes to defend ASV systems against adversarial attacks using a separate detection network. A VGG-like network is introduced to determine whether an input is a genuine or an adversarial sample. Our method is demonstrated to be effective on detecting adversarial samples. We also analyze various kinds of unseen attack setting's impact on detector robustness. 
We observe that the detector is relatively robust against substitute ASV systems, while the generalizability against perturbation methods is not symmetric and detector performance could drop greatly in some cases. The weak robustness against unseen perturbation methods shows a direction for developing stronger countermeasures. 

%% file: acknowledgements.tex
\section{Acknowledgements}
This work is partially supported by HKSAR Government's Research Grants Council General Research Fund (Project No. 14208718).

%% file: main.bbl
\begin{thebibliography}{10}
\providecommand{\url}[1]{#1}
\csname url@samestyle\endcsname
\providecommand{\newblock}{\relax}
\providecommand{\bibinfo}[2]{#2}
\providecommand{\BIBentrySTDinterwordspacing}{\spaceskip=0pt\relax}
\providecommand{\BIBentryALTinterwordstretchfactor}{4}
\providecommand{\BIBentryALTinterwordspacing}{\spaceskip=\fontdimen2\font plus
\BIBentryALTinterwordstretchfactor\fontdimen3\font minus
  \fontdimen4\font\relax}
\providecommand{\BIBforeignlanguage}[2]{{%
\expandafter\ifx\csname l@#1\endcsname\relax
\typeout{** WARNING: IEEEtran.bst: No hyphenation pattern has been}%
\typeout{** loaded for the language `#1'. Using the pattern for}%
\typeout{** the default language instead.}%
\else
\language=\csname l@#1\endcsname
\fi
#2}}
\providecommand{\BIBdecl}{\relax}
\BIBdecl

\bibitem{kinnunen2012vulnerability}
T.~Kinnunen, Z.~Wu, K.~Lee, F.~Sedlak, E.~Chng, and H.~Li, ``Vulnerability of
  speaker verification systems against voice conversion spoofing attacks: The
  case of telephone speech,'' in \emph{ICASSP}.\hskip 1em plus 0.5em minus
  0.4em\relax IEEE, 2012, pp. 4401--4404.

\bibitem{shchemelinin2014vulnerability}
V.~Shchemelinin, M.~Topchina, and K.~Simonchik, ``Vulnerability of voice
  verification systems to spoofing attacks by {TTS} voices based on
  automatically labeled telephone speech,'' in \emph{International Conference
  on Speech and Computer}.\hskip 1em plus 0.5em minus 0.4em\relax Springer,
  2014, pp. 475--481.

\bibitem{wu2015spoofing}
Z.~Wu, N.~Evans, T.~Kinnunen, J.~Yamagishi, F.~Alegre, and H.~Li, ``Spoofing
  and countermeasures for speaker verification: A survey,'' \emph{speech
  communication}, vol.~66, pp. 130--153, 2015.

\bibitem{wu2015asvspoof}
Z.~Wu, T.~Kinnunen, N.~Evans, J.~Yamagishi, C.~Hanil{\c{c}}i, M.~Sahidullah,
  and A.~Sizov, ``Asvspoof 2015: the first automatic speaker verification
  spoofing and countermeasures challenge,'' in \emph{Interspeech}, 2015.

\bibitem{kinnunen2017asvspoof}
T.~Kinnunen, M.~Sahidullah, H.~Delgado, M.~Todisco, N.~Evans, J.~Yamagishi, and
  K.~A. Lee, ``The asvspoof 2017 challenge: Assessing the limits of replay
  spoofing attack detection,'' in \emph{Interspeech}, 2017.

\bibitem{todisco2019asvspoof}
M.~Todisco, X.~Wang, V.~Vestman, M.~Sahidullah, H.~Delgado, A.~Nautsch,
  J.~Yamagishi, N.~Evans, T.~Kinnunen, and K.~A. Lee, ``Asvspoof 2019: Future
  horizons in spoofed and fake audio detection,'' in \emph{Interspeech}, 2019.

\bibitem{williams2019speech}
J.~Williams and J.~Rownicka, ``Speech replay detection with x-vector attack
  embeddings and spectral features,'' in \emph{Interspeech}, 2019.

\bibitem{cai2019dku}
W.~Cai, H.~Wu, D.~Cai, and M.~Li, ``The {DKU} replay detection system for the
  asvspoof 2019 challenge: On data augmentation, feature representation,
  classification, and fusion,'' \emph{arXiv preprint arXiv:1907.02663}, 2019.

\bibitem{hanilcci2015classifiers}
C.~Hanil{\c{c}}i, T.~Kinnunen, M.~Sahidullah, and A.~Sizov, ``Classifiers for
  synthetic speech detection: A comparison,'' in \emph{Interspeech}, 2015.

\bibitem{wu2012detecting}
Z.~Wu, E.~S. Chng, and H.~Li, ``Detecting converted speech and natural speech
  for anti-spoofing attack in speaker recognition,'' in \emph{Interspeech},
  2012.

\bibitem{correia2014preventing}
M.~J. Correia, A.~Abad, and I.~Trancoso, ``Preventing converted speech spoofing
  attacks in speaker verification,'' in \emph{International Convention on
  Information and Communication Technology, Electronics and Microelectronics
  (MIPRO)}.\hskip 1em plus 0.5em minus 0.4em\relax IEEE, 2014, pp. 1320--1325.

\bibitem{kreuk2018fooling}
F.~Kreuk, Y.~Adi, M.~Cisse, and J.~Keshet, ``Fooling end-to-end speaker
  verification with adversarial examples,'' in \emph{ICASSP}.\hskip 1em plus
  0.5em minus 0.4em\relax IEEE, 2018, pp. 1962--1966.

\bibitem{li2020adversarial}
X.~Li, J.~Zhong, X.~Wu, J.~Yu, X.~Liu, and H.~Meng, ``Adversarial attacks on
  {GMM} i-vector based speaker verification systems,'' in \emph{ICASSP}.\hskip
  1em plus 0.5em minus 0.4em\relax IEEE, 2020, pp. 6579--6583.

\bibitem{chen2019real}
G.~Chen, S.~Chen, L.~Fan, X.~Du, Z.~Zhao, F.~Song, and Y.~Liu, ``Who is real
  bob? adversarial attacks on speaker recognition systems,'' \emph{arXiv
  preprint arXiv:1911.01840}, 2019.

\bibitem{li2020practical}
Z.~Li, C.~Shi, Y.~Xie, J.~Liu, B.~Yuan, and Y.~Chen, ``Practical adversarial
  attacks against speaker recognition systems,'' in \emph{International
  Workshop on Mobile Computing Systems and Applications}, 2020, pp. 9--14.

\bibitem{xie2020real}
Y.~Xie, C.~Shi, Z.~Li, J.~Liu, Y.~Chen, and B.~Yuan, ``Real-time, universal,
  and robust adversarial attacks against speaker recognition systems,''
  \emph{arXiv preprint arXiv:2003.02301}, 2020.

\bibitem{liu2019adversarial}
S.~Liu, H.~Wu, H.-y. Lee, and H.~Meng, ``Adversarial attacks on spoofing
  countermeasures of automatic speaker verification,'' \emph{arXiv preprint
  arXiv:1910.08716}, 2019.

\bibitem{kurakin2016adversarial}
A.~Kurakin, I.~Goodfellow, and S.~Bengio, ``Adversarial machine learning at
  scale,'' \emph{arXiv preprint arXiv:1611.01236}, 2016.

\bibitem{song2017pixeldefend}
Y.~Song, T.~Kim, S.~Nowozin, S.~Ermon, and N.~Kushman, ``Pixeldefend:
  Leveraging generative models to understand and defend against adversarial
  examples,'' \emph{arXiv preprint arXiv:1710.10766}, 2017.

\bibitem{xu2017feature}
W.~Xu, D.~Evans, and Y.~Qi, ``Feature squeezing: Detecting adversarial examples
  in deep neural networks,'' \emph{arXiv preprint arXiv:1704.01155}, 2017.

\bibitem{gong2017adversarial}
Z.~Gong, W.~Wang, and W.-S. Ku, ``Adversarial and clean data are not twins,''
  \emph{arXiv preprint arXiv:1704.04960}, 2017.

\bibitem{wang2019adversarial}
Q.~Wang, P.~Guo, S.~Sun, L.~Xie, and J.~H. Hansen, ``Adversarial regularization
  for end-to-end robust speaker verification,'' \emph{Proc. Interspeech 2019},
  pp. 4010--4014, 2019.

\bibitem{wu2020defense}
H.~Wu, S.~Liu, H.~Meng, and H.-y. Lee, ``Defense against adversarial attacks on
  spoofing countermeasures of asv,'' in \emph{ICASSP}, 2020.

\bibitem{goodfellow2014explaining}
I.~Goodfellow, J.~Shlens, and C.~Szegedy, ``Explaining and harnessing
  adversarial examples,'' \emph{arXiv preprint arXiv:1412.6572}, 2014.

\bibitem{samizade2019adversarial}
S.~Samizade, Z.-H. Tan, C.~Shen, and X.~Guan, ``Adversarial example detection
  by classification for deep speech recognition,'' \emph{arXiv preprint
  arXiv:1910.10013}, 2019.

\bibitem{nagrani2020voxceleb}
A.~Nagrani, J.~S. Chung, W.~Xie, and A.~Zisserman, ``Voxceleb: Large-scale
  speaker verification in the wild,'' \emph{Computer Speech \& Language},
  vol.~60, p. 101027, 2020.

\bibitem{dehak2010front}
N.~Dehak, P.~Kenny, R.~Dehak, P.~Dumouchel, and P.~Ouellet, ``Front-end factor
  analysis for speaker verification,'' \emph{IEEE Transactions on Audio,
  Speech, and Language Processing}, vol.~19, no.~4, pp. 788--798, 2010.

\bibitem{snyder2018x}
D.~Snyder, D.~Garcia-Romero, G.~Sell, D.~Povey, and S.~Khudanpur, ``X-vectors:
  Robust {DNN} embeddings for speaker recognition,'' in \emph{ICASSP}.\hskip
  1em plus 0.5em minus 0.4em\relax IEEE, 2018, pp. 5329--5333.

\bibitem{zeinali2019but}
H.~Zeinali, S.~Wang, A.~Silnova, P.~Mat{\v{e}}jka, and O.~Plchot, ``{BUT}
  system description to {V}oxceleb speaker recognition challenge 2019,''
  \emph{arXiv preprint arXiv:1910.12592}, 2019.

\bibitem{papernot2016limitations}
N.~Papernot, P.~McDaniel, S.~Jha, M.~Fredrikson, Z.~B. Celik, and A.~Swami,
  ``The limitations of deep learning in adversarial settings,'' in \emph{IEEE
  European symposium on security and privacy (EuroS\&P)}.\hskip 1em plus 0.5em
  minus 0.4em\relax IEEE, 2016, pp. 372--387.

\bibitem{xiang2019margin}
X.~Xiang, S.~Wang, H.~Huang, Y.~Qian, and K.~Yu, ``Margin matters: Towards more
  discriminative deep neural network embeddings for speaker recognition,''
  \emph{arXiv preprint arXiv:1906.07317}, 2019.

\bibitem{nagrani2017voxceleb}
A.~Nagrani, J.~S. Chung, and A.~Zisserman, ``Voxceleb: a large-scale speaker
  identification dataset,'' \emph{arXiv preprint arXiv:1706.08612}, 2017.

\bibitem{kingma2014adam}
D.~P. Kingma and J.~Ba, ``Adam: A method for stochastic optimization,''
  \emph{ICLR}, 2015.

\end{thebibliography}
